\documentstyle[11pt,aaspp4]{article}

\slugcomment{Submitted to the Astron. J.}

\lefthead{Levshakov, Tytler \& Burles}
\righthead{Deuterium to Hydrogen towards QSO 1009+2956} 

\begin{document}

\title{Deuterium to hydrogen towards QSO 1009+2956
from a mesoturbulent model\altaffilmark{1}}

\author{Sergei A. Levshakov,\hspace{0.3cm}David Tytler}
\affil{National Astronomical Observatory, Mitaka, Tokyo 181, Japan}

\and

\author{Scott Burles}
\affil{University of Chicago, 5640 S. Ellis Ave., Chicago, IL 60637}

\altaffiltext{1}{Based on observations obtained at the W. M. Keck Observatory,
which is jointly operated by the University of California and the California Institute
of Technology.}

\begin{abstract}
We present a new analysis of the deuterium absorption at $z = 2.504$ towards the 
quasar Q1009+2956 
using the mesoturbulent model which accounts for possible correlations in
the large scale velocity field. 
We obtain a slightly higher deuterium-to-hydrogen ratio 
D/H $\simeq (3.5 - 5.0)\times10^{-5}$ as compared with the 
recent measurement D/H = $(3.3 - 4.5)\times10^{-5}$ (68\% C.L.) performed by
Burles \& Tytler (1998b) using the usual
microturbulent approximation which assumes that the velocity field is 
uncorrelated.
Other mesoturbulent calculations of the D-abundances 
at $z = 3.572$ towards Q1937--1009 
and at $z = 0.701$ towards Q1718+4807 
(the systems showing `low' and possibly `high' D/H values, respectively,
in the microturbulent approaches)
agree  with the present one within the errors of measurements.
Thus, the mesoturbulent analysis does not reveal any spatial variations of D/H and 
supports the standard homogeneous model of big bang nucleosynthesis.

\end{abstract}


\section{Introduction}

High resolution spectra of the H+D absorption in Lyman
limit systems towards distant quasars and the subsequent Voigt profile
fitting analysis [VPF] suggested a dispersion in 
the primordial hydrogen isotopic ratio D/H 
$\equiv$ N(\ion{D}{1})/N(\ion{H}{1})
(the ratio of the \ion{D}{1} and \ion{H}{1} column densities)
of about one order of magnitude (see a short summary in Burles \& Tytler 1998b). 
This finding provoked a lively discussion in the literature
on inhomogeneous models of big bang nucleosynthesis (e.g., Jedamzik \& Fuller 1995;
Webb {\it et al.} 1997; Kurki-Suonio {\it et al.} 1997).
The standard (homogeneous) BBN predicts
the same D/H abundance ratios in different directions in the early Universe
since `no realistic astrophysical process other than the Big Bang could
produce significant D' (Schramm 1998, p.6).
Deuterium is created exclusively in BBN and therefore
we can expect that the D/H ratio decreases with cosmic time due to conversion 
of D into $^3$He and heavier elements in stars. 
It is clear that the precise measurements
of the D/H values at high redshift are extremely important to probe whether BBN
was homogeneous.
The choice of the appropriate BBN model
may in turn place constrains on different
models of structure formation. 

The lack of information about the nature of the 
absorption line broadening mechanism means that we might obtain different
values of D/H if we make different assumptions.
The multicomponent microturbulent models which are normally used assume that
there is no correlation in the large scale velocity field. 
In these models
both thermal and non-thermal velocities along the sight-line are represented
by plain Gaussian distributions and the total velocity dispersion which is
measured in high resolution spectra ($b_{obs}$) is given by
$b^2_{obs} = b^2_{therm} + b^2_{turb}$.
However, if there are correlations in the velocity field, the 
velocity distribution along a given line of sight may deviated 
significantly from the Gaussian model,  i.e. 
$b^2_{obs} \neq b^2_{therm} + b^2_{turb}$.
The application of the standard VPF analysis to such systems may then
give incorrect D/H.
We do not know, however, how to distinguish 
between the multicomponent microturbulent and 
one component mesoturbulent models if both
of them show  good fits to the spectral data. 

To decide which model is better additional observations 
can be considered. For instance,
if we assume that the Lyman limit systems arise in the outer regions of 
intervening galactic halos, then direct observations of the large scale flows
in the giant Ly$\alpha$ emission halos at $z > 2$ (van Ojik {\it et al.} 1997),
or in the star forming galaxies at $z \simeq 3$ (Pettini {\it et al.} 1998),
or the complex metal absorption-line features
in the quasar absorber/galaxy pairs at $z \sim 1$ (Bergeron {\it et al.} 1992;
Bechtold \& Ellinfson 1992; Charlton \& Churchill 1998)
may favor the mesoturbulent approximation since the rms turbulent velocity
($\sigma_t \equiv b_{turb}/\sqrt{2}$)
in these halos ($\sigma_t \sim 20 - 100$~km~s$^{-1}$)
is larger that the thermal width ($v_{th} \equiv b_{therm}$)
of the hydrogen lines
($v_{th} \sim 13 - 15$~km~s$^{-1}$). The correlation effects become important
if $\sigma_t/v_{th} \gtrsim 1$. Examples are given in 
Levshakov {\it et al.} (1997).

The first D/H measurement at $z = 2.504$ towards Q1009+2956 was performed by 
Tytler \& Burles (1997). They found D/H in the range from $2.2\times10^{-5}$
to $4.2\times10^{-5}$. Later Burles \& Tytler (1998b, hereafter BT) reconsidered
this measurement including full coverage of the Lyman series and better quality
data above and below the Lyman continuum break. Besides BT used the new method  
to measure D/H in quasar absorption systems described in Burles \& Tytler (1998a).
Considering six different microturbulent models, BT found 
D/H = $(3.9 \pm 0.6)\times10^{-5}$
which is consistent with their results on 
the $z = 3.572$ system towards Q1937--1009 where
D/H = $(3.3 \pm 0.3)\times10^{-5}$ (Burles \& Tytler 1998a).
Neither measurement is consistent with the D/H at 
$z = 0.701$ towards Q1718+4807 deduced using the microturbulent model
with a single component: 
D/H = $(2.0 \pm 0.5)\times10^{-4}$ (Webb {\it et al.} 1997), and
D/H = $8\times10^{-5} < {\rm D/H} < 57\times10^{-5}$
(Tytler {\it et al.} 1998).
This discord can be resolved using a more complex velocity model for 
Q1718+4807. Tytler et al. (1998) considered a second velocity component with
the microturbulent model, while Levshakov et al. (1998) used a
single component mesoturbulent model.

Here we apply the mesoturbulent model to Q1009+2956, and we conclude that
all three QSOs may be described by
a single value of D/H which is approximately equal to $4\times10^{-5}$.

\section{The measurement of D/H}

The present study is primarily aimed at the inverse problem in the analysis of
the H+D Ly$\alpha$ absorption. The higher order Lyman series lines observed by BT
(Ly$\beta$, Ly-6, Ly-12, Ly-13, Ly-14, Ly-20, and Ly-21)
were also chosen in this analysis
to restrict the set of possible velocity field configurations.
We consider an absorbing region of thickness $L$ which is a continuous medium
(presumably the outer region of a foreground galaxy). The absorber is assumed to
exhibit a mixture of bulk motions such as infall and outflows, tidal flows etc.
Then the motion along the sight-line may be characterized by a fluctuating velocity
field that we consider as a continuous random function of the space coordinate, $v(s)$.

For simplicity, we assume a homogeneous density $n_e$ and temperature $T_{kin}$.
We have only weak justifications for these assumptions, which are
necessitated by the complexity of the analysis.
The assumption of constant $T_{kin}$ may be supported by the fact that 
the states of thermally
stable, optically thin interstellar gas take only restricted temperature ranges 
over a relatively wide range of density variations (see, e.g., Donahue \& Shull 
1991). The constant $n_e$ seems to be quite plausible approximation for this 
particular $z = 2.504$ system since observational data show that metal lines 
from different ions show similar profiles.
In Section~3 we use the similarity of 
\ion{C}{4} and \ion{Si}{4} profiles to estimate $T_{kin}$ independently on the
analysis of hydrogen lines.
But these arguments do not rule out possible variations in $T_{kin}$
and/or $n_e$ in general.
Our present mesoturbulent model is similar to that adopted in Levshakov {\it et al.} (1998a,b).

We have developed a computational procedure which allows us
to estimate physical parameters and simultaneously an appropriate velocity field
structure along the sight-line. The details of the computational scheme 
based on the reverse Monte Carlo technique [RMC]
are given in Levshakov {\it et al.} (1999). 
The algorithm requires to define a simulation box for the 
parameter vector $\theta = \{$ D/H, N(\ion{H}{1}), $T_{\rm kin}$,
$\sigma_{\rm t}/v_{\rm th}, L/l \}$;
here $l$ is the velocity field correlation length. The
continuous random function  $v(s)$ is represented by
its sampled values at equal space intervals $\Delta s$, i.e. by
the sequence $v_j$ ($j = 1, 2, \dots , k)$
of the velocity components parallel to the line of sight
at the spatial points $s_j$. We describe $v_j$ by 
a two-point Markovian process.

In the present study we adopt for the physical parameters the following boundaries~:
N(\ion{H}{1}) ranges from
$2.0\times10^{17}$ to $3.0\times10^{17}$ cm$^{-2}$, -- the total hydrogen column
density measured from the Lyman continuum optical depth by BT was found to lie
in the interval $(2.1 - 2.8)\times10^{17}$~cm$^{-2}$;
D/H -- from
$3.0\times10^{-5}$ to $3.0\times10^{-4}$;
$T_{\rm kin}$ -- from $10^4$ to $2.5\times10^4$~K.
For $\sigma_{\rm t}/v_{\rm th}$ the boundaries were set from 1.0 to 2.5 
to be typical for that observed in galactic halos. Since for
$L/l \gg 1$ the meso- and microturbulent profiles tend
to be identical (Levshakov \& Kegel 1997),
we consider here only moderate $L/l$ ratios in the range 1~--~5.
We do not assume that the D and H lines have identical velocities to
the metal lines. The \ion{H}{1} distribution is fixed by the
higher order Lyman series lines.
We fix $z = 2.503571$ (the value adopted by BT)
as the reference radial velocity at which $v = 0$.

We use a $\chi^2$ minimization routine and the RMC optimization technique to
estimate five model parameters and the appropriate configuration of the
velocity field $v(s)$. Note that $k$ components of the velocity field here are 
the `nuisance parameters'  and their number does not influence the
physical parameters we are interested in.
A lower bound to $k$ is given by the inequality 
[cf. eq.(22) in Levshakov {\it et al.} 1997]~: 
\begin{equation} 
\frac{L}{\Delta s} > \frac{L}{l}\,\frac{2}{
\ln\,\left[1 - (v_{\rm th}/\sigma_{\rm t})^2\right]^{-1}}\; ,
\label{eq:E1} 
\end{equation} 
with $\sigma_{\rm t}/v_{\rm th} > 1$.

In the Q1009+2956 spectrum, there are many absorption features
blending the Lyman series lines. We do not fit these additional
absorptions since they do not affect significantly the D/H
measurement in this system, as shown by BT. The strongest
absorption feature seen at $z = 2.50456$ in Fig.~1 has, according to BT,
N(\ion{H}{1}) = $3.7\times10^{13}$ cm$^{-2}$,
which is less than 0.02\% of the total N(\ion{H}{1}) at $z = 2.504$.

The portions of the Lyman series lines that, after preliminary analysis,
were chosen as most appropriate to the simultaneous RMC fitting are
indicated in Fig.~1 by the thick gray lines. In the standard 
$\chi^2$ minimization, the following objective function is usually used
\begin{equation} 
{\cal L} \equiv \chi^2 = \frac{1}{\nu}\,
\sum^n_{j=1}\,\sum^{m_j}_{i=1}\,\frac{\left[ I_i - F(\lambda_i,\theta) \right]^2}
{\sigma^2_{I_i}}\; .
\label{eq:E2} 
\end{equation} 
Here, $I_i$ and $\sigma_{I_i}$ are the observed normalized intensity and
the experimental error within the $i$th pixel of the line profile, respectively.
$F(\lambda_i,\theta)$ is the simulated intensity at the same $i$th
pixel having wavelength $\lambda_i$. The total number of hydrogen lines is
labeled by $n$, and the total number of data points 
$m = \sum^n_{j=1} m_j$, where $m_j$ is the number of data points for the
$j$th line. The number of fitted parameters $p = 5$, and $\nu = m - p$ is
the degree of freedom ($\nu = 208$ in our case).

Equation (\ref{eq:E2}) assumes that all $\lambda_i$-values are known
exactly, and thus it ignores errors in the wavelength calibrations
($\lambda$-errors).
These errors can be neglected in case of not very steep profiles and
not very high SNR data, when small errors in $\lambda$ leave 
the intensity within its uncertainty range, i.e.
$$ 
I + \left| \frac{dI}{d\lambda} \right|\sigma_\lambda < I + \sigma_I\: ,
$$ 
or
\begin{equation} 
\left| \frac{dI}{d\lambda} \right| <
\frac{\sigma_I}{\sigma_\lambda}\; ,  
\label{eq:E3} 
\end{equation} 
where $\sigma_\lambda = \left(\sigma^2_{stat} + \sigma^2_{system}\right)^{1/2}$
is the error of the wavelength scale calibration, and $\sigma_I$ is the
experimental uncertainty in $I$.

If, however, inequality~(\ref{eq:E3}) is violated, then the calibration errors,
appearing in the shift of the line profile, lead to inappropriate high values
of ${\cal L}$. To correct such solutions, we should adjust the 
$\lambda$ scale.
We have calculated~(\ref{eq:E3}) for different lines under consideration (see Table~1). 
From this table it is seen that (\ref{eq:E3}) is strongly violated for 
the Ly$\alpha$, Ly$\beta$ and Ly-6 lines, but is fulfilled
for the Ly-12 and higher order hydrogen lines.

To account for the $\lambda$-adjustment 
we incorporate new fitting parameters $\delta\lambda$
in equation (\ref{eq:E2}) which now gets the form
\begin{equation} 
\chi^2 = \frac{1}{\nu'}\left\{
\sum^n_{j=1}\,\sum^{m_j}_{i=1}\,\frac{\left[ 
I_i - F(\lambda_i+\delta\lambda_j,\theta) \right]^2}
{\sigma^2_{I_i}} + 
\sum^{n'}_{j=1}\,\left( 
\frac{\delta\lambda_j}{\sigma_{\lambda_j}} \right)^2 \right\}\; ,
\label{eq:E4} 
\end{equation} 
where the correction factor $\delta\lambda_j$ is common for all
pixels of the $j$th hydrogen line. $n'$ is the number of lines whose
wavelength scales are to be corrected ($0 \leq n' \leq n$), and
$\sigma_{\lambda_j} \equiv \sigma_\lambda \simeq \frac{1}{4} {\rm FWHM}$.
The number of fitted parameters in this case is $p' = 5 + n'$
and hence $\nu' = m - p'$.
In this approach possible values of $\delta\lambda_j$ are restricted
and lie in the range $| \delta\lambda_j | < \sigma_\lambda$.

Using the original continuum as determined by BT 
(without allowing for errors in the local continuum in 
the vicinity of each hydrogen line) 
we found a few RMC profile fits which are listed in Table~2.
We tried to find solutions with the reduced
$\chi^2$ per degree of freedom of $\chi^2_{min} \lesssim 1$.
For all models from Table~2 the calculated hydrogen profiles
are similar to that shown in Fig.~1.
We do not find any large differences between calculated and real spectra.
The under-absorption on the blue-ward side of D Ly$\alpha$ is probably caused
by the Ly$\alpha$ forest lines. 
We do not treat this part of the spectrum because, in this case,
unsaturated higher order Lyman series lines provide accurate predictions
for the velocity of the deuterium Ly$\alpha$ line and its shape.
Besides, as shown by BT, 
these additional absorptions do not affect significantly the D/H ratio
in the $z = 2.504$ system.

The derived $v(s)$ configurations are not unique. But their projections -- 
the radial-velocity distribution functions $p(v)$  -- 
are very much alike (see examples in Fig.~2).
Figure~2 shows the $p(v)$ distributions 
found by the RMC procedure (histograms) and for comparison  
$p(v)$ for the best three-component microturbulent model 2 found by BT 
({\it thin-line} Gaussians and their weighted sum shown by {\it dotted-line}).
Both the RMC $p(v)$ and the combined BT $p(v)$
distributions show similar asymmetry.
But the interpretation of these $p(v)$ is different. In the microturbulent
model the asymmetry is caused by the individual clouds with different physical parameters,
whereas the mesoturbulent solution describes the homogeneously distributed gas with
constant density and temperature along the sight-line.

Now comes the question about the most probable kinetic temperature 
and the accuracy of the approximation $n_e$~= constant.
Table~2 shows the spread of the $T_{kin}$ values between 13700~K and 18100~K
for different models. Each of these models is acceptable from the statistical
point of view. To select the adequate models additional observational data
should be taken into account. For this purpose we consider below high SNR data
on the unsaturated \ion{C}{4} and \ion{Si}{4} lines observed at the same
$z = 2.504$ towards Q1009+2956.

\section{The measurement of $T_{kin}$ from metal absorption lines}

We turn now to a brief discussion of the application of the Entropy-Regularized
$\chi^2$-Minimization [ERM] procedure 
developed by Levshakov {\it et al.} (1998c, hereinafter LTA) 
to recover independently the kinetic temperature from the \ion{C}{4} and
\ion{Si}{4} data obtained by BT. We suppose that \ion{H}{1} and metals from 
the same Lyman limit system should give similar $T_{kin}$.

The ERM procedure utilizes complex but similar absorption line profiles of
different ions to estimate a single value of $T_{kin}$ for the whole absorbing
region. The similarity of the complex profiles of ions with different masses
and ionization potentials stems from the homogeneous gas density distribution
along the sight-line, otherwise we would expect to observe different intensity
fluctuations within the line profiles caused by variations of the local ionization
parameter $U$ which is the ratio of the density of ionizing photons to the
total gas density. 
We consider the complex structure of the absorption lines as being generated
by the large scale motions with the correlated internal structure.
Satisfactory ERM solutions ($\chi^2_{min} \lesssim 1$) for a pair
of lines of different ions will only exist when the
homogeneous density approximation is consistent with the spectra.

The absorber in question is the outer part of the halo of a distant galaxy.
The gas in the halo photo-ionized by QSOs is optically thin in the Lyman continuum.
For such system the equilibrium temperature is weakly dependent on $U$
(Donahue \& Shull 1991). We may expect, therefore, that a single $T_{kin}$
is still a suitable approximation in this case.

To estimate $T_{kin}$ for the particular $z = 2.504$ system, we have
chosen \ion{C}{4}$\lambda$1548 (SNR = 72) and \ion{Si}{4}$\lambda$1394 (SNR = 73)
lines observed with 8~km~s$^{-1}$ spectral resolution (FWHM) by BT. 
We did not use the red components of these doublets because their data are
more noisy. Applying the ERM procedure to the \ion{C}{4}$\lambda$1548 
and \ion{Si}{4}$\lambda$1394 lines, we obtained the gas temperature
$T^\ast_{kin} \simeq 13500$~K. 

Figures~3 and 4 illustrate our results. 
The best-fitting profiles with three equidispersion components 
[for definition see eq.(12) in LTA]
are shown
in panels $a$ and $b$ of Fig.~3 by solid lines with the tick marks indicating
the positions for the separate components. The choice of the
appropriate $T^\ast_{kin} \simeq 13500$~K corresponding to the optimal value of the 
normalized regularization parameter $\hat{\alpha}_{opt} = 0.174$ is shown in Fig.~4.
The objective function used in this case has the form
\begin{equation}
{\cal L}_\alpha = \chi^2 + \alpha\,\psi\; ,
\label{eq:E5} 
\end{equation} 
where $\psi$ is a penalty function [eq.(21) in LTA] 
and the regularization parameter $\alpha$ is given by
\begin{equation}
\alpha = \alpha_{min} + \hat{\alpha}\,(\alpha_{max} - \alpha_{min})\; .
\label{eq:E6} 
\end{equation} 
For this particular case $\alpha_{min} = - 0.19$ and $\alpha_{max} = 0.05$ were chosen. 

The obtained ERM estimation of $T^\ast_{kin} \simeq 13500$~K
lies in the range of $T_{kin}$ found by the RMC method from the H+D Ly$\alpha$
and higher order Lyman series profiles. 
Since the accuracy of the ERM solutions is shown to be about 10\% (LTA), 
models {\bf a}, {\bf e}, and {\bf g} in Table~2 seem to be consistent with
the ERM result.

\section{Conclusion}

We have shown that the mesoturbulent model can explain the data adequately.
In the following we discuss the relationship between the  
micro- and mesoturbulence models, and related issues.

The mesoturbulent model assumes constant $T_{kin}$ and $n_e$ but allows velocity
correlations, while the microturbulent model allows different but constant
$T_{kin}$ and $n_e$ for each of a finite number of components and assumes no
coherent structure in the velocity field.

In the microturbulent approach it is explicitly assumed that the radial
velocities are normally distributed so that the line profile is every time
uniquely determined. In contrast, in the mesoturbulent approach we do not
make the assumption of a Gaussian distribution for the radial velocities.
This leads in turn to the concepts of randomness and unpredictability in the
line formation process. That is to say,
the apparent $b$-parameters usually measured in high resolution spectra may not
be suitable to determine the true ratio $\sigma_t/v_{th}$ which is used
to estimate the relative importance of the non-thermal broadening
(see Levshakov \& Kegel 1997, for details).

In the two approaches, the D/H ratios 
are found to be similar for the particular
$z = 2.504$ system. However, microturbulent models yield slightly lower deuterium
abundance and rather narrow range ($\simeq$ 15\%) of its uncertainty. In the
mesoturbulent solutions, although
the total hydrogen column densities lie inside
the range found by BT, the D/H values are less restricted because, in general,
the radial-velocity distribution function $p(v)$ is not known {\it a priori}.
Our analysis gives a certain range for 
N(\ion{H}{1})$_{\rm tot}$ and D/H due to the finite signal-to-noise ratio
and different $p(v)$ configurations.
This is shown in Fig.~5
where different confidence regions  for
this pair of physical parameters are depicted for models {\bf a},  {\bf e}, and {\bf g}
under the assumption that the other parameters
$T_{\rm kin}$, $\sigma_{\rm t}/v_{\rm th}$, $L/l$ and $v(s)$ are
fixed (the computing procedure is described in Levshakov {et al.} 1999).
The estimated D-abundance towards Q1009+2956 
[D/H $\simeq (3.5 - 5.0)\times10^{-5}$]
agrees within the errors of measurement with the RMC results 
for the \ion{D}{1} absorption systems seen
towards Q1937-1009 
[D/H $\simeq (3.8 - 4.8)\times10^{-5}$~: Levshakov {\it et al.} 1998a] and
towards Q1718+4807 
[D/H $\simeq (3.0 - 7.5)\times10^{-5}$~: Levshakov {\it et al.} 1998b, Levshakov 1998].

In general, accurate measurement of D/H requires ($i$) correct modeling,
($ii$) treatment of continuum placement errors reported by Burles \& Tytler
(1998a,b), and ($iii$) treatment of contaminating absorption.

We conclude that the current observations
support SBBN and that there is no conflict with
the D/H measurements in these three Lyman limit systems.
A single and robust value of D/H~$\simeq 4\times10^{-5}$ 
is sufficient to describe H+D profiles within the framework of the
mesoturbulent model. 
 
\acknowledgments

S.A.L. and D.T. gratefully acknowledge the hospitality of the National
Astronomical Observatory of Japan where this work was performed.

\newpage

\figcaption[]{
Observations (normalized flux) and RMC fits for Q1009+2956.
The Keck/HIRES echelle data obtained by BT --
dots and 1$\sigma$ error bars; the calculated RMC profiles (model {\bf a} in Table~2)
convolved with the instrumental resolution of FWHM = 8~km~s$^{-1}$ -- solid curves.
The thicker grey parts of these curves show the regions about each
Lyman line used in the fitting procedure.
The redshift of this Lyman limit system is $z = 2.503571$, 
according to BT. 
\label{fig1}
}
 
\figcaption[]{
Radial-velocity distribution functions $p(v)$  
of the RMC solutions for models {\bf a}, {\bf e}, and {\bf g} 
(solid, dashed  and dotted line histograms, respectively) and
for comparison three $p(v)$ distributions
with the velocity dispersions $b({\rm H})$ = 16.0, 16.8,
and 24.6 km~s$^{-1}$ (solid curves)
adopted from the best fit microturbulent model~2 of BT.
The curves are weighted by the corresponding
hydrogen column densities of
$1.29\times10^{17}, 0.50\times10^{17}$ and $0.24\times10^{17}$ cm$^{-2}$.
The sum of all three weighted functions is shown by the dotted curve.
\label{fig2}
}

\figcaption[]{
($a,b$) Normalized data points with $1\sigma$ error bars for HIRES echelle
spectrograph observations (FWHM = 8~km~s$^{-1}$) on the Keck telescope of the
\ion{C}{4}$\lambda1548$ (SNR = 72) and \ion{Si}{4}$\lambda1394$ (SNR = 73)
at the redshift $z = 2.503571$ towards the quasar Q1009+2956 (BT). The ERM
result (reduced $\chi^2 = 0.92$ with 37 degrees of freedom) for \ion{C}{4}
and \ion{Si}{4} (solid lines) is based on simultaneously fitting the multiple
equidispersion lines located at the positions shown by the tick marks at the
top of each panel. The evaluated kinetic temperature $T^\ast \simeq 13500$~K
is plotted in panel $a$. 
\label{fig3}
}

\figcaption[]{
An example of the Entropy-Regularized $\chi^2$-Minimization [ERM] technique used to
measure the kinetic temperature from the complex absorption spectra of the 
\ion{C}{4} and \ion{Si}{4} lines shown in Fig.~3. The filled circles connected by
dotted line show the normalized values of the cross-entropy $\hat{\cal K}$ as function
of the normalized regularization parameter $\hat{\alpha}$.
The dashed curve is the normalized curvature of the $\hat{\cal K}(\hat{\alpha})$
trajectory. Its maximum at point $\hat{\alpha} = 0.174$ corresponds to the kinetic
temperature $T^\ast \simeq 13500$~K which is indicated by arrows.
For more details (especially on the ERM technique) the reader is encouraged to consult LTA.
\label{fig4}
}

\figcaption[]{
Confidence regions (68\% confidence level)
in the plane `N(\ion{H}{1})--D/H' for solutions {\bf a}, 
{\bf e}, and {\bf g} listed in Table~2.
Different solutions give different contours.
The other parameters of models {\bf a}, {\bf e}, and {\bf g} --
$T_{\rm kin}$, $\sigma_{\rm t}/v_{\rm th}$, and $L/l$ and
the corresponding configurations of the velocity fields $v(s)$ are fixed. 
The size of the confidence region depends on signal to noise.
The letters $a, e,$ and $g$ mark the
points of maximum likelihood for models~{\bf a}, {\bf e}, and {\bf g} (see Table~2).
\label{fig5}
}

\newpage

\begin{deluxetable}{ccccc}
\footnotesize
\tablecaption{\ion{H}{1} data for the $\lambda$-adjustment criterion
(eq.(\ref{eq:E3})). 
\label{tab1}}
\tablewidth{16cm}
\tablehead{
\colhead{ } & 
\colhead{Ly$\alpha$} &
\colhead{Ly$\beta$} &
\colhead{Ly-6} &
\colhead{Ly-12} 
}
\startdata
$\lambda\lambda$, \AA & 4258.30 - & 3593.01 - & 3260.50 - & 3213.03 -\nl
                      & 4258.50 & 3593.21 & 3260.66 & 3213.21\nl
$\langle {\rm SNR} \rangle$ & 80 & 36 & 17 & 13\nl
$\sigma_I/\sigma_\lambda$, \AA$^{-1}$ & 0.4 & 1.0 & 3.0 & 4.0\nl
$dI/d\lambda$, \AA$^{-1}$ & 3.0 & 4.0 & 5.0 & 4.0\nl
\enddata
\end{deluxetable}

\begin{deluxetable}{cccccccccc}
\footnotesize
\tablecaption{Model parameters derived from the Lyman series lines
by the RMC method\tablenotemark{a}. 
\label{tab2}}
\tablewidth{16cm}
\tablehead{
\colhead{Model} & \colhead{D/H} &
\colhead{N$_{17}$} &
\colhead{$T_4$} &
\colhead{$\sigma_{\rm t}/v_{\rm th}$} &
\colhead{$L/l$} & \colhead{$\chi^2_{min}$} & 
\colhead{$\delta\lambda_\alpha$} &
\colhead{$\delta\lambda_\beta$} &
\colhead{$\delta\lambda_6$} 
}
\startdata
({\bf a}) &3.76&2.37&1.38&1.75&2.03&0.80&0.018&$-0.019$&$-0.022$\nl
({\bf b}) &4.26&2.17&1.81&1.18&2.18&0.84&0.028&$-0.014$&$-0.022$\nl
({\bf c}) &4.34&2.20&1.57&2.00&1.68&0.73&0.021&$-0.013$&$-0.022$\nl
({\bf d}) &4.37&2.20&1.55&1.58&2.17&0.73&0.026&$-0.011$&$-0.022$\nl
({\bf e}) &4.39&2.16&1.51&1.58&1.65&0.72&0.023&$-0.014$&$-0.022$\nl
({\bf f}) &4.51&2.19&1.64&1.09&2.03&0.83&0.023&$-0.015$&$-0.022$\nl
({\bf g}) &4.65&2.15&1.37&1.17&3.70&0.71&0.026&$-0.011$&$-0.022$\nl
\enddata
\tablenotetext{a}{
${\rm N}_{17}$ is the total hydrogen column density in
units of $10^{17}$ cm$^{-2}$, D/H in units of $10^{-5}$,
$T_4$ kinetic temperature in units of $10^4$ K,
$\delta\lambda_j$ adjustment factor in \AA\ , where 
$\delta\lambda_\alpha$, $\delta\lambda_\beta$, and $\delta\lambda_6$
are the wavelength corrections for
Ly$\alpha$, Ly$\beta$, and Ly-6, respectively. 
}
\end{deluxetable}

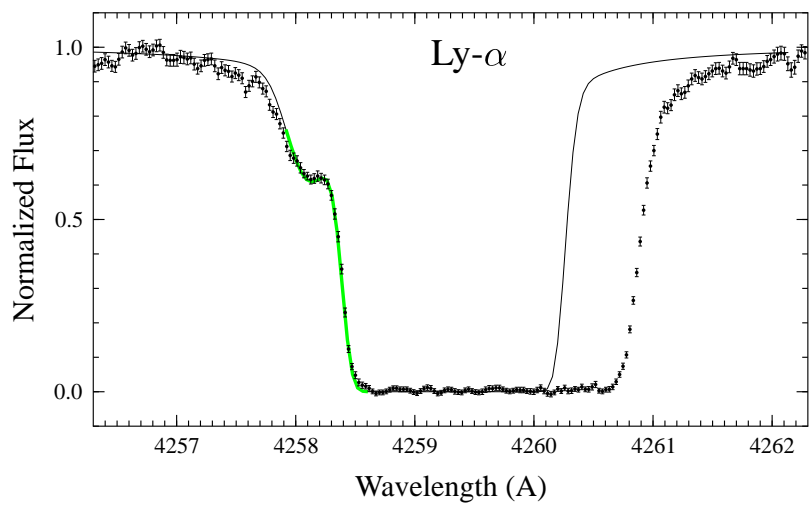
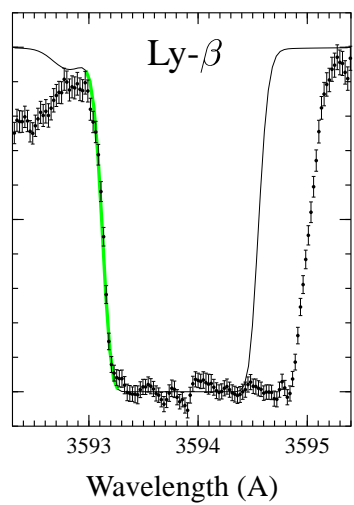
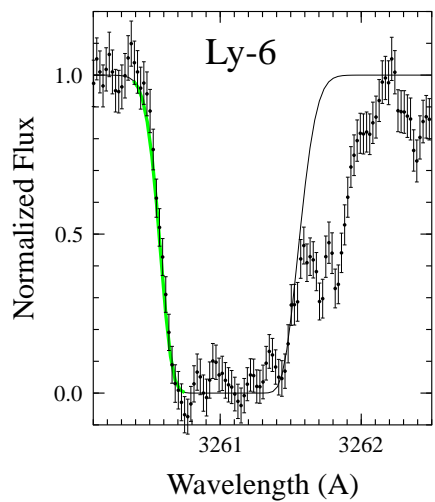
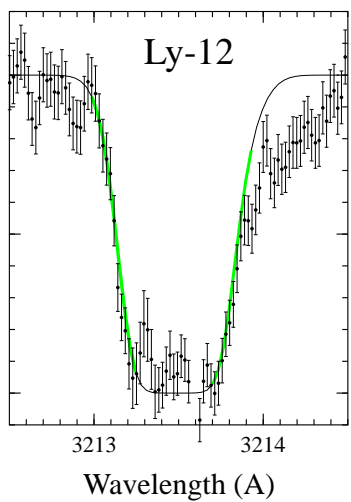
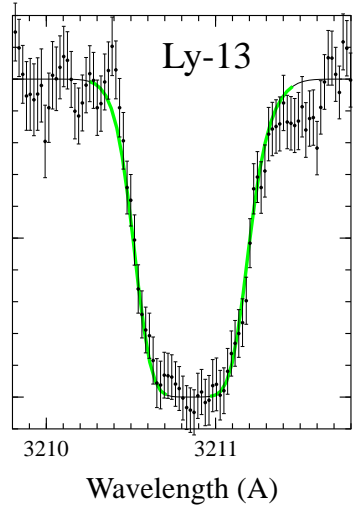
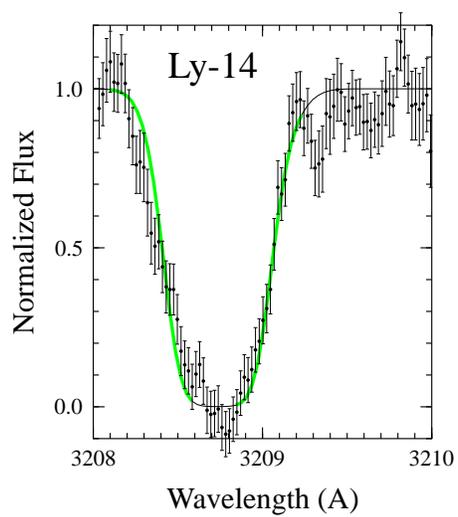
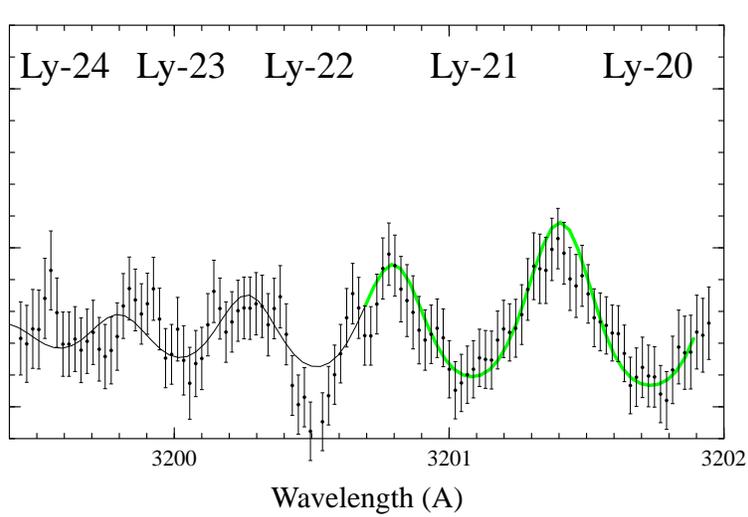
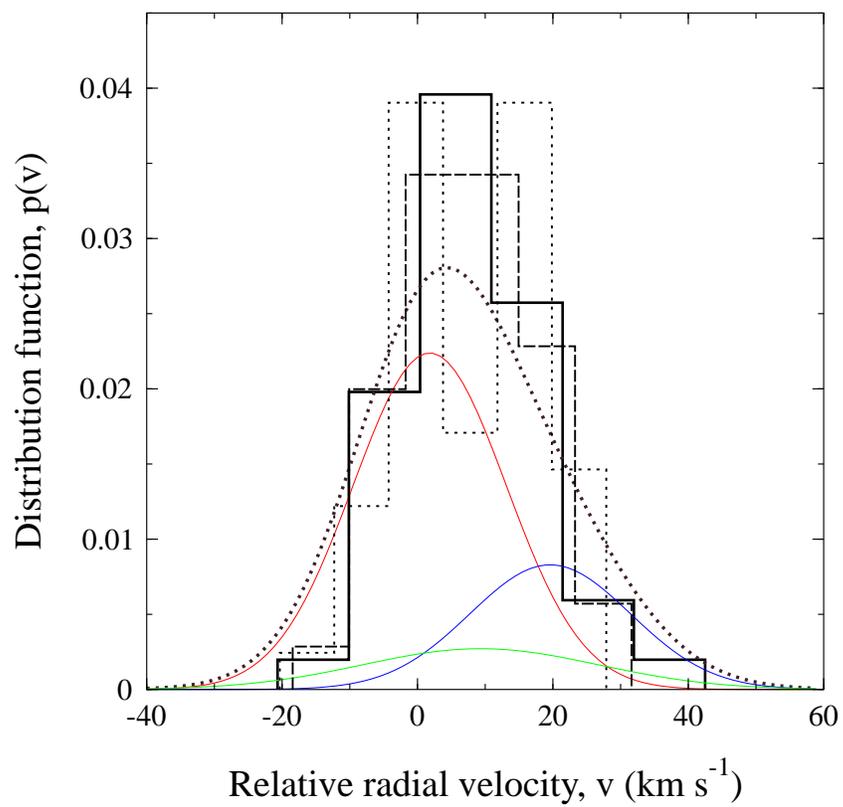
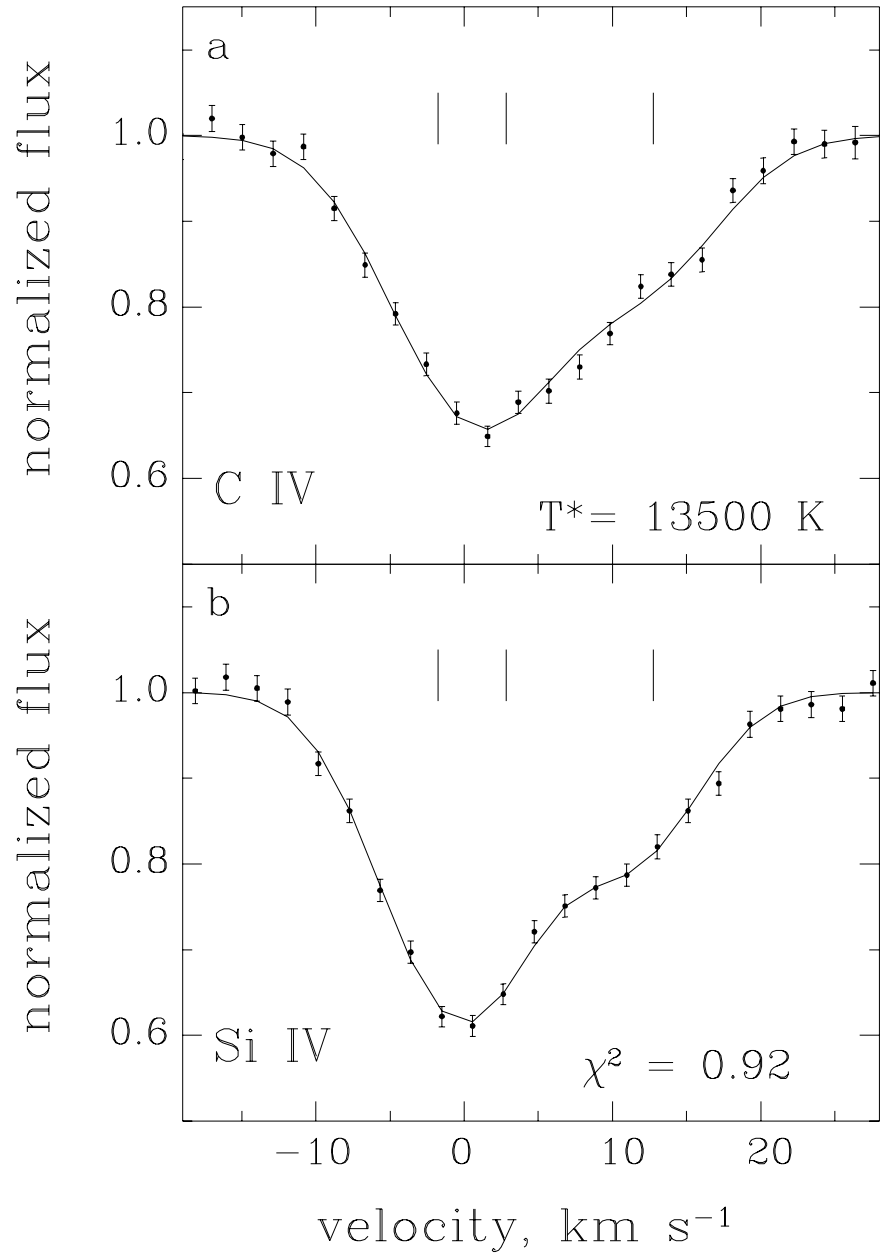
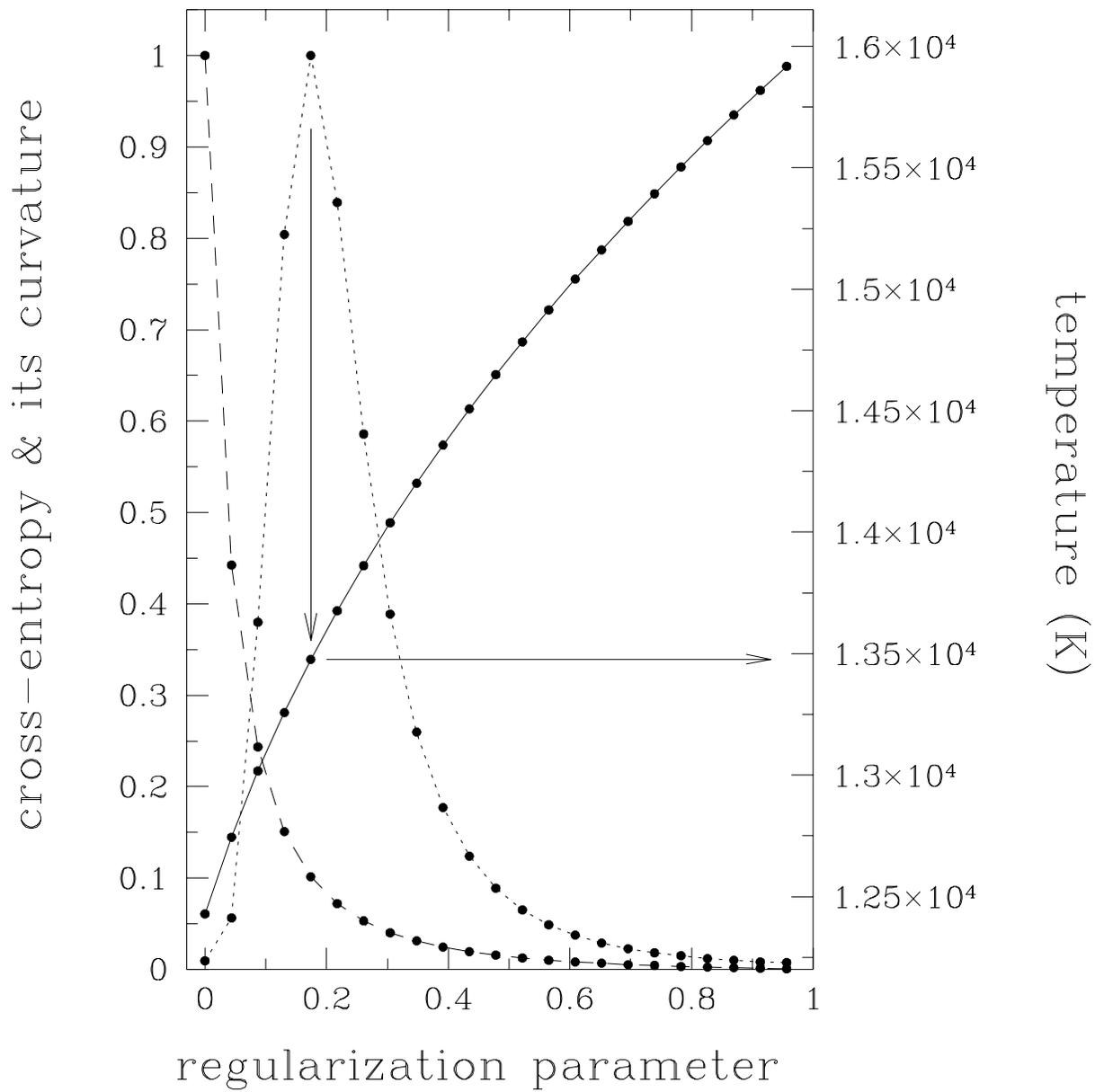
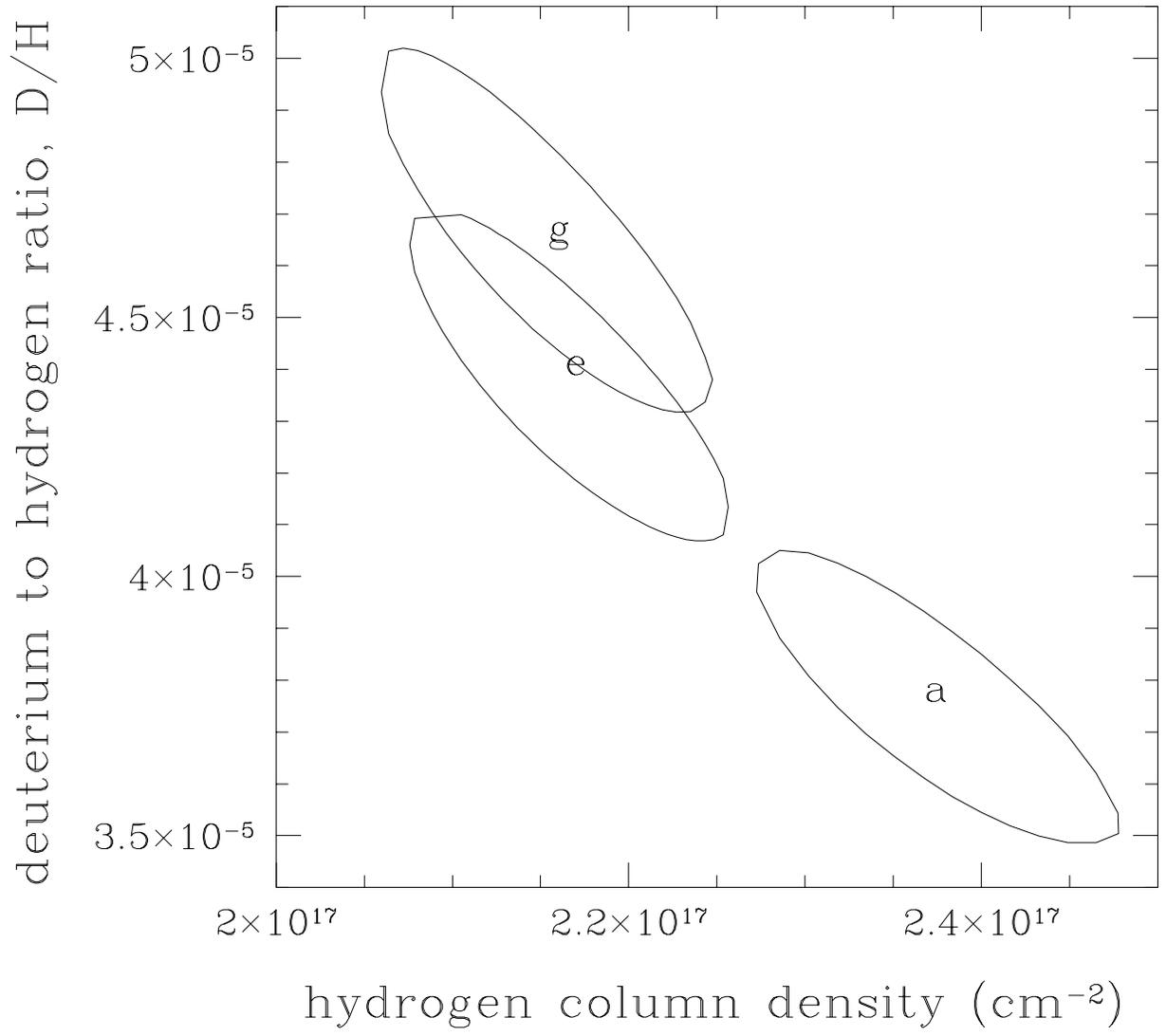
\end{document}